\documentclass[letterpaper]{usenix}
\usepackage[noadjust]{cite}
\usepackage{graphicx}
\usepackage{subfigure}
\usepackage{url}

\usepackage[delayed, graphics, dvips, showlabels,sections,textmath,displaymath]{preview}

\newcommand{\beq}{\begin{equation}}
\newcommand{\eeq}{\end{equation}}
\def\bearn{\begin{eqnarray*}}
\def\eearn{\end{eqnarray*}}
\def\barr{\begin{array}}
\def\earr{\end{array}} 

\def\bt{BitTorrent}

\newcommand{\urlsamefont}[1]
{
\urlstyle{same}\url{#1}
}

\begin{document}
\sloppy

\title{De-anonymizing BitTorrent Users on Tor}

\author{
\authname{Stevens Le Blond\thanks{Poster NSDI'10, San Jose, CA, April
    28-30, 2010. The two first authors contributed
    equally to this work}, Pere Manils, Abdelberi Chaabane}
\and
\authname{Mohamed Ali Kaafar, Arnaud Legout, Claude Castellucia, Walid
Dabbous}
\authaddr{I.N.R.I.A}
}

\maketitle 
\sloppy

Privacy of users in Peer-to-peer (P2P) networks goes far beyond their
current usage and is a fundamental requirement to the adoption of P2P
protocols for legal usage. In a climate of cold war between P2P
filesharing users and anti-piracy groups, more and more users are
moving to anonymizing networks in an attempt to hide their
identity. However, when not designed to protect users information, a
P2P protocol often leaks information that compromises the identity of
its users.

\bt{} is a P2P filesharing protocol that is daily used by millions of
users but that has not been designed to protect the anonymity of its
users. Indeed, it has recently been shown that an adversary can
continuously spy, i.e., collect the IP-to-contents mapping, on most
\bt{} users of the Internet and from a single machine
\cite{angling}. In addition to spy on BitTorrent users, an
attacker might be able to exploit \bt{} control messages to
de-anonymize a user behind an anonymizing network such as Tor.

Tor relies on onion routing over an overlay network maintained by
volunteers to anonymize TCP applications such as web browsing, P2P
filesharing, etc. To reach the Internet via Tor, an application
selects $3$ Tor nodes at random and then first encrypts its messages
with the key shared with the last node (exit node), then with the key
of the $2$nd node, and finally with the key of the $1$st node. The $3$
Tor nodes that route a user's messages form a \textit{circuit} and all
TCP \textit{streams} created by that user during a 10-minutes period
will be multiplexed into one, or a few circuits. Each Tor node in a
circuit then decrypts/encrypts the messages after routing them to/from
the Internet. Onion routing thus guarantees that no Tor node knows
both the source IP address and the payload of a message.

A BitTorrent user may use Tor to (1) connect to a server (tracker) to
collect lists of peers sharing a file, (2) connect to other peers to
distribute a file, or (3) both.

In this proposal, we instrument $6$ exit nodes for a period of $23$
days to demonstrate that an attacker can de-anonymize \bt{} users for
any of the $3$ aforementioned usages by volunteering to maintain an
exit node and eavesdropping appropriate BitTorrent control
messages. In addition, as all streams are multiplexed into the same
circuit, we show that de-anonymizing one \bt{} stream allows to
potentially de-anonymize other applications such as web browsing. We
propose $3$ attacks to de-anonymize BitTorrent users on Tor.

Our first attack consists in inspecting the payload of some \bt{}
control messages and search for the public IP address of a user. In
particular, the \textit{announce} messages that a client sends to the
tracker to collect a list of peers distributing a content, and the
\textit{extended handshake} messages that some clients send right
after the application handshake sometimes contain the public IP
address of the user. However, we have not tested the accuracy of the
public IP address contained in those messages so we do not consider
them in the following.

Our second attack consists in rewriting the list of peers returned by
the tracker to include the IP address of a peer that we control. As
the user will then \textit{directly} connect to the peer controlled by
the attacker, the latter can de-anonymize the user by inspecting the
IP header. Whereas this hijacking attack is accurate, it only works
when the user relies on Tor only to connect to the tracker.

Finally, the third attack consists in exploiting the DHT to search for
the public IP address of a user. Indeed, whereas Tor does not support
UDP, BitTorrent's DHT uses UDP for transport and when a BitTorrent
client fails to contact the DHT using its Tor interface, it reverts to
its public interface hence publishing its public IP address into the
DHT. As the content identifier and the port number of a client transit
through the exit node, and port numbers are uniformly distributed, an
attacker can use this information to identify a BitTorrent user in the
DHT. This DHT attack is very accurate and works even when the peer
uses Tor to connect to other peers.

Using the hijacking and DHT attacks, we de-anonymized and profiled
close to $9,000$ public IP addresses of BitTorrent users on Tor. In
particular, we have exploited the multiplexing of streams from
different applications into the same circuit to profile the web
browsing habits of the BitTorrent users on Tor.

\begin{small}

\end{small}

\end{document}